\documentclass{svproc}
\usepackage{url}
\usepackage{bm}
\usepackage{graphicx}
\usepackage{multicol}
\usepackage{amsmath,amssymb,amsfonts}
\usepackage{textcomp}
\usepackage{multirow}

\begin{document}
\mainmatter              
\title{A Research and Strategy of Remote Sensing Image Denoising Algorithms}
\titlerunning{Image Denoising}  
%
\author{Ling Li\inst{1 2}\and Junxing Hu\inst{1 2} \and Fengge Wu\inst{2} \and Junsuo Zhao\inst{2}}
\authorrunning{Ling Li et al.} 
%
%
\institute{University of Chinese Academy of Sciences, Beijing, China,\\
\email{liling2017@iscas.ac.cn}
\and
Institute of Software Chinese Academy of Sciences (ISCAS),\\
Beijing, China}

\maketitle              

\begin{abstract}
Most raw data download from satellites are useless, resulting in transmission waste, one solution is to process data directly on satellites, then only transmit the processed results to the ground. Image processing is the main data processing on satellites, in this paper, we focus on image denoising which is the basic image processing. There are many high-performance denoising approaches at present, however, most of them rely on advanced computing resources or rich images on the ground. Considering the limited computing resources of satellites and the characteristics of remote sensing images, we do some research on these high-performance ground image denoising approaches and compare them in simulation experiments to analyze whether they are suitable for satellites. According to the analysis results, we propose two feasible image denoising strategies for satellites based on satellite TianZhi-1.
\end{abstract}
\section{Introduction}
We want to find an approach suitable for image denoising on satellites. The study of remote sensing images has always been a hot subject, because large amounts of valuable information can be obtained by the efficient use of remote sensing images. However, if all images acquired by a satellite are directly transmitted to the ground, most of them are found useless, and this may cause waste of data transmission, resulting in transmission pressure, thus, image processing on satellites is required, and only transmit the useful data to the ground, image denoising is the premise of other image processing, such as object detection, semantic segmentation.

The difference between image denoising on the ground and image denoising on satellites can be analyzed from two aspects: 1) Image noises can be caused by internal disturbances such as mechanical motion, mechanical materials, and internal equipment circuits, as well as external disturbances such as electromagnetic waves, celestial discharges, etc. The imaging on the ground can largely avoid causing noises, but many external disturbances in imaging on satellites can not be avoid, such as the illumination, the atmosphere, the disturbing photographic object distance and the high-velocity and unpredictable rotation of cosmic materials, which all make the camera on a satellite hard to acquire a legible image or image sequence of the same scene that includes the enough information we want. 2) Compared with image denoising on the ground, there is no advanced computing resources for image denoising on satellites, and few clean remote sensing images can be used to clearn.


In this paper, we analyze the high-performance approaches of image denoising on the ground from different angles, compare the adaptability and feasibility for image denoising on satellites, and find a suitable approach or a potential approach for image denoising on satellites. To the best of our knowledge, this is the first paper to combine traditional approaches and deep learning approaches for experimental comparison to find a strategy suitable for satellites.


The rest of the paper is organized as follows. Section 2 details some outstanding image denoising approaches. The various simulation results of these approaches are compared in Section 3. And in Section 4, we propose two image denoising strategies for satellites. Finally, Section 5 concludes this paper.

\section{Image Denoising Approaches}
Image denoising is the most basic inverse problem in image processing. The noisy image can be formulate as
\begin{equation}
\bm{y=x+n}
\end{equation}
where \textbf{\emph{y}} is the corrupt signal, \textbf{\emph{x}} is the original signal, and \textbf{\emph{n}} is the noisy signal.

\begin{figure}
\centering
\includegraphics[width=8cm]{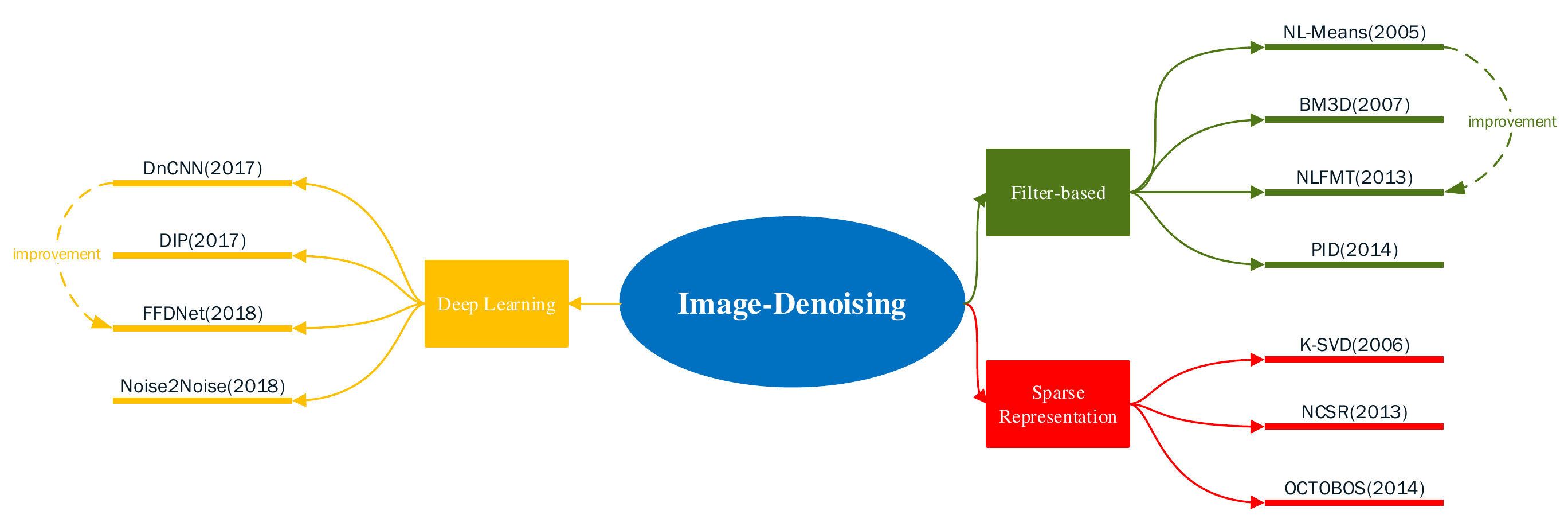}
\caption{All approaches that appeared in this paper.}
\label{fig:1}
\end{figure}

In this section, we divide the current outstanding image denoising approaches into three categories (see Fig.\ref{fig:1}): (1) Filter-based denoising approaches are relatively traditional, they process images by designing filters. (2) Natural images have sparse representation under a certain mode, while random noises cannot be sparse. We classify such approaches using sparsity to constrain images into sparse representation approaches. (3) Deep learning approaches are very popular image processing approaches currently, most of them summarize the intrinsic properties of the noisy images with other similar but no-noise images, and exploit the external images to create a priori condition and then denoise the noisy image.
\subsection{Filter-based Approaches}
The filter-based approaches are traditional approaches with strong theoretical support. We list four breakthrough filter-based approaches in this subsection. Image denoising filters can operate directly on the image itself (spatial domain) or in its frequency domain. The spatial filter finds out the relationship between the image pixels to remove noises, and the remaining complex noises can be removed by transforming the image into frequency domain.
\subsubsection{NL-means Filter and its Method noise Thresholding using Wavelets (NLFMT).}
NL-means \cite{bua:col:mor} is an early breakthrough approach, unlike the previous useing image local information to denoise image, it uses the entire image redundant information, and achieves a better result. NLFMT \cite{kum} aims to remedy the image details which be wrongly removed in NL-means, relieve the smoothness of denoising images, and gives a better visual perception.
\subsubsection{Block-Matching and 3D filtering (BM3D)} \cite{dab:foi:kat:egi} is similar to NL-means, it finds similar blocks in the image, too, but it is more complicated. It not only integrates spatial method and transform method but also takes advantage of both intra-fragment correlation and inter-correlation. Although it was proposed in 2012, it still can achieve state-of-the-art denoising results so far.
\subsubsection{Progressive Image Denoising (PID)} simplifies image denoising process and improves the visible artifacts. Knaus et al. \cite{kna:zwi:pid} extended the work of dual-domain image denoising (DDID) \cite{kna:zwi:ddid},and regarded image denoising as a simple physical process which progressively reduces noises by deterministic annealing (DA).
%
\subsection{Sparse Representation Approaches}
In this subsection we describe how to denoise images using sparse representation and show different ways to use it in different approaches. Given a patch vector $\bm{x}_k\in\mathbb{R}^{n}$ of an image $\bm{x}$ and a dictionary $\bm{\Phi}\in\mathbb{R}^{n{\times}M}$, $M$ is the number of atoms in $\bm{\Phi}$. The spare coding process of $\bm{x}_k$ over $\bm{\Phi}$ is to find a sparse vector $\bm{\alpha}_k\in\mathbb{R}^{M}$ such that $\bm{x}_k{\approx}\bm{\Phi}\bm{\alpha}_k$. Then the entire image can be sparsely represented by the set of sparse codes $\{\bm{\alpha}_k\}$.
\subsubsection{K-SVD} uses singular value decomposition (SVD) to update the dictionary for K iterations \cite{ela:aha}. And the dictionary can separately trained on patches of the noisy images and high-quality images.

\subsubsection{Nonlocally Centralized Sparse Representation (NCSR)} \cite{don:zha:shi:li} uses the image nonlocal self-similar to obtain the coefficient of sparse coding for evaluating the original image, and changed the goal of image noise reduction to suppress the sparse coding noise.

\subsubsection{OCTOBOS} \cite{wen:rav:bre} learns a collection of well-conditioned square transforms \cite{rav:bre} and allows groups of patches with common features to better sparsified by their own texture-specific transform.
%
\subsection{Deep Learning Approaches}
Compared with the previous two categories, deep learning approaches can exploit parallel architectures to build more complicated computational models and achieve similar or better results for image denoising. In 2009, Jain et al. \cite{jai:seu} applied convolution network to denoise images, though the effect is general, since then, more scholars began to research deep learning in image denoising. In the early deep image denoising models, such as, MLP \cite{bur:sch:har}, CSF \cite{sch:rot} and TNRD \cite{che:poc}, a specific model is trained for a certain noise level, limiting the wide use of the approach. For practical reasons, we only describe representative approaches that can handle multilevel noises with a single model, proposed in recent years.
%
\subsubsection{DnCNN} \cite{zha:zuo:che:men:zha} uses the residual learning strategy to implicitly remove the clean image in the hidden layers of the network, then return the denoised image by the noisy image minus the network output.
\subsubsection{Deep Image Prior (DIP)} \cite{uly:ved:lem} restricts the number of iterations in the parameter learning process to get the clean image by exploiting the parametrization performs low impedance to clean signals and high impedance to noises.
\subsubsection{FFDNet} \cite{zha:zuo:zha} is an improved network for DnCNN, its architecture contains a DnCNN in the middle. FFDNet operates on downsampled sub-images, while DnCNN operates directly on noisy images.

\subsubsection{Noise2Noise} applies the basic statistics to the signal reconstruction of machine learning \cite{leh:mun:has:lai:kar:ait:ail}. It can learn the latent clean image $\bm{x}$ from the corrupted image $\bm{y}$ by getting the smallest average deviation according to loss function $\bm{L}$:
\begin{equation}
\arg\min_{\bm{x}}\mathbb{E}_{\bm{y}}\{\bm{L}(\bm{x},\bm{y})\}
\end{equation}
\section{Simulation Comparison of Representative Approaches}
In this section, we simulate these above mentioned image denoising approaches by using the official source codes to conduct comparative analysis, we mainly compare the results in PSNR, SSIM, time cost and memory consumption (or model parameters). To do this, we use a machine with 64Gb RAM, Intel(R) Xeon(R) E5-2680 @v4 2.40GHz processor and one Nvidia TITAN Xp GPU card.
\subsection{Dataset}
\begin{figure}
\centering
\includegraphics[width=6cm]{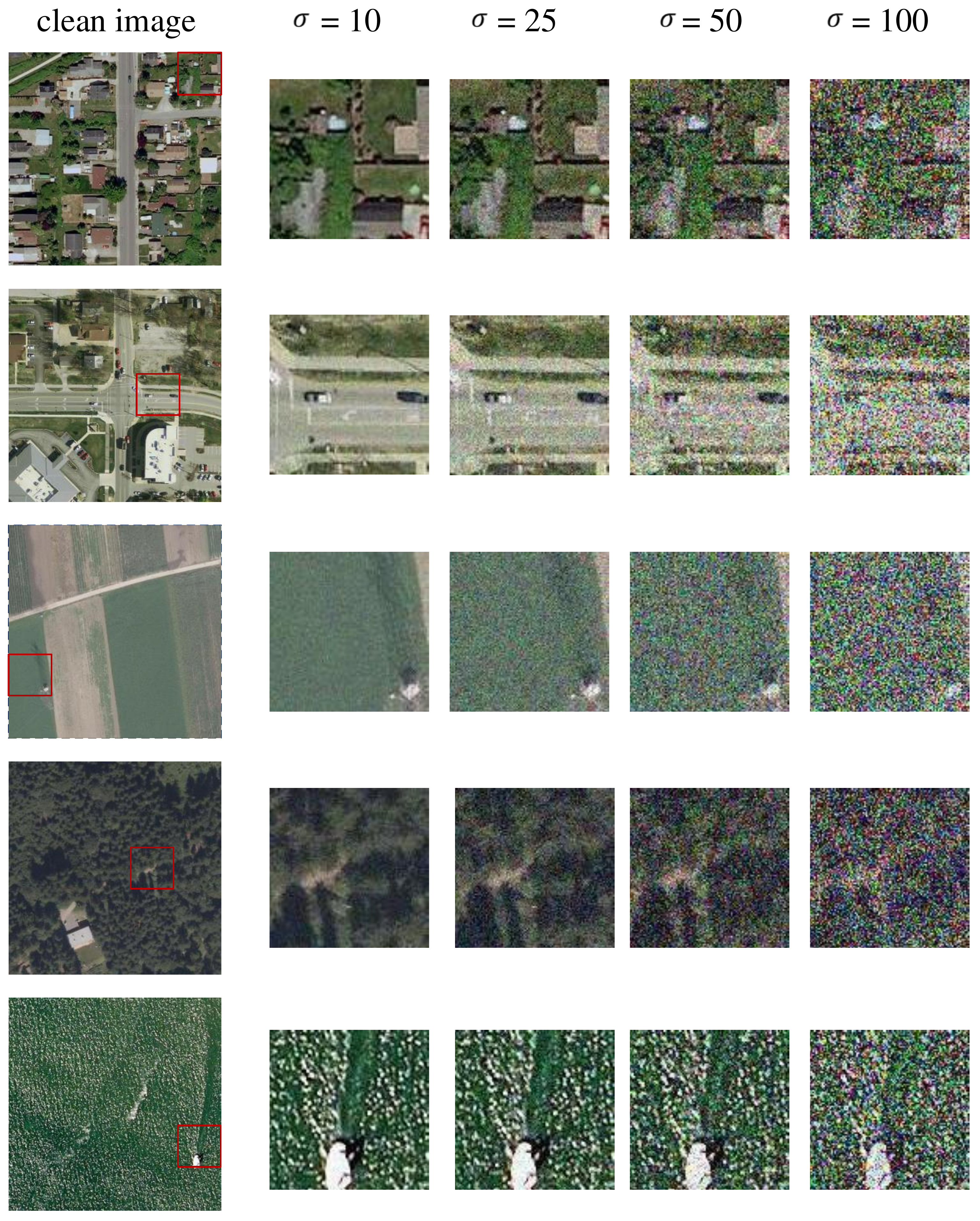}
\caption{The clean test images and image details of the four noise levels.}
\label{fig:datatest}
\end{figure}

We apply the Inria Aerial Image dataset \cite{mag:tar:cha:all} to our simulation experiments, these images have a resolution of 0.3 meters and cover residential areas, roads, grasslands, forests, lakes, etc (see Fig.\ref{fig:datatest}). We compress the original TIF image format into JPEG format, and cut the images into size $512\times512$, then randomly select 792 images from the cut images as the training set for deep learning approaches, and select another 5 images respectively contains residential areas, roads, grasslands, forests, and lakes as the test set. We add Additive White Gaussian Noise (AWGN) with one of the four noise levels ($\sigma=10,25,50$ and 100) to each clean image in train set randomly but add each noise level to each image in test set, i.e., each test image corresponds to four noisy images (see Fig.\ref{fig:datatest}). The images with noise level $\sigma=100$ are tested as outliers.
%
%
\subsection{PSNR and SSIM}
\begin{figure}
\centering
\includegraphics[width=9cm]{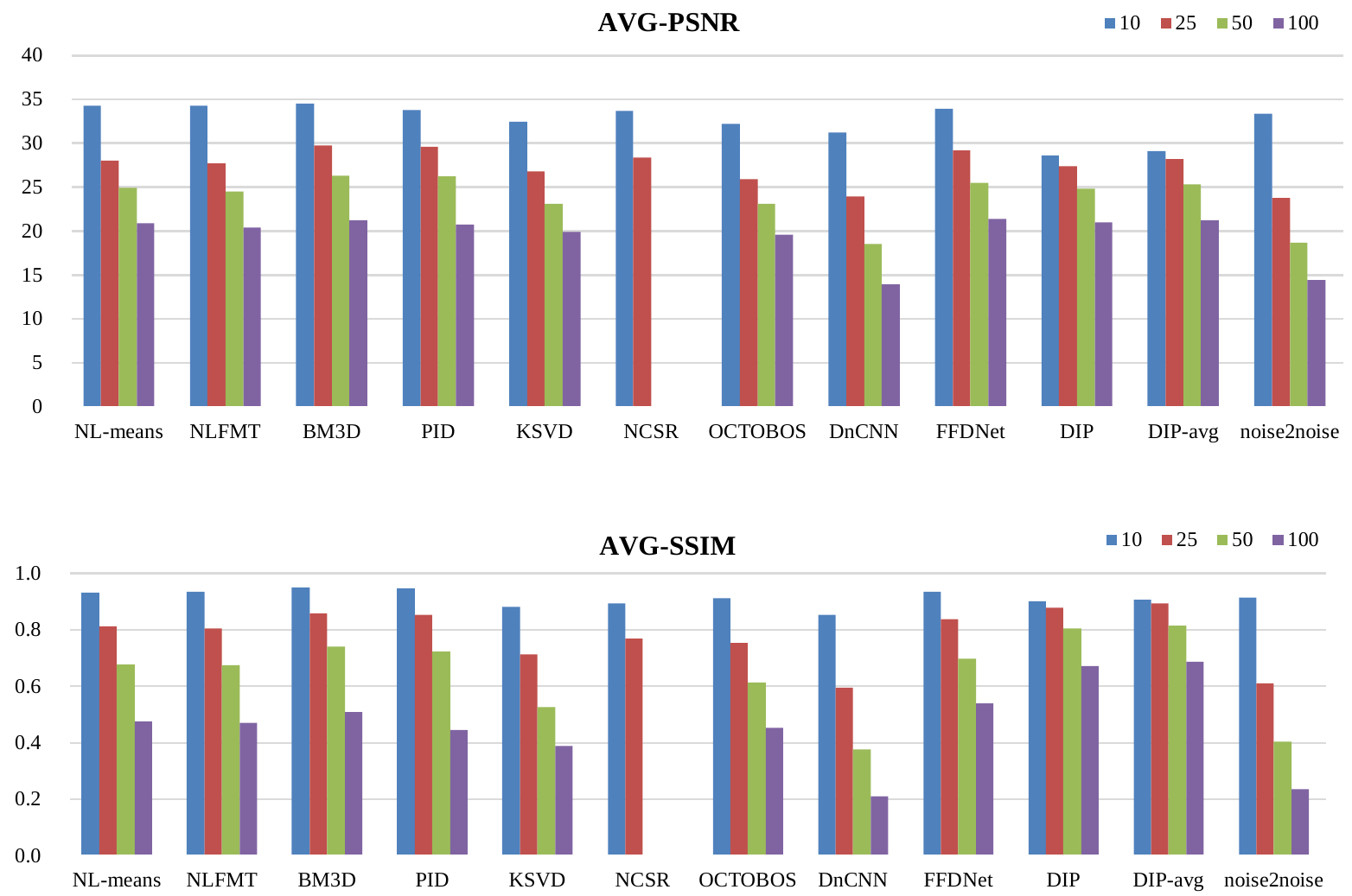}
\caption{The average PSNR and average SSIM of test images at each noise level.}
\label{fig:psnr_ssim}
\end{figure}

To compare the image denoising qualities of the above mentioned approaches, we choose Peak Signal to Noise Ratio (PSNR) and Structural Similarity Index (SSIM) to evaluate the quality of a processed image. The larger the value, the better the denoised image quality. Fig.\ref{fig:psnr_ssim} shows the average PSNR and average SSIM for each approach on denoising five images at each specific noise level.

From Fig.\ref{fig:psnr_ssim}, we found filter-based approaches have similar and pretty image denoising qualities, the average PSNR is slightly higher than the other two categories, and so is SSIM while the noise level is low. The image denoising qualities of sparse representation approaches are just a little inferior to those of the filter-based approaches, and with the increase of the noise level, our images can not complete the noise reduction using NCSR, failing in pixel intensities K-mens clustering. The image denoising qualities of deep learning approaches are uneven, PSNR and SSIM of FFDNet are comparable to those of filter-based approaches, DIP and DIP-avg only perform well on high proportion noises, DnCNN and noise2noise have higher PSNR and SSIM in low proportion noises, but not good under high proportion noises.

\subsection{Time Cost and Memory Consumption}
Time cost and memory consumption are key indicators that can be used to evaluate whether an approach is applicable to satellites. For deep learning approaches, we only care about the performance of the trained models, not about training models, except DIP, due to DIP directly denoises the noisy image by generating clean image without training the network.
\subsubsection{Time Cost.} We record the image denoising time for images with a noise level of $\sigma=25$. Table 1 shows the image denoising time on CPU of filter-based approaches and sparse representation approaches. It is easy to find that BM3D is much faster in image denoising, approximately 3 seconds, compared to the other approaches. For deep learning approaches, they have an advantage in the time of image denoising, only take less than 0.1 seconds on GPU and several seconds on CPU when using trained models to denoise an image, except DIP which is about 1,000 seconds on GPU.
\begin{table}
\caption{The time cost of image denoising using filter-based approaches and sparse representation approaches on CPU.}
\begin{center}
\begin{tabular}{l@{\quad}l@{\quad}l@{\quad}l@{\quad}l@{\quad}l@{\quad}l}
\hline
\rule{0pt}{12pt}CPU\_time& Residential area& Lake& Road& Grassland& Forest& Average\\[2pt]
\hline
\rule{-3pt}{12pt}
NL-means  &     1,657.30& 1,653.90& 1,655.90& 1,657.10& 1,657.20& 1,656.28\\
NLFMT  &  1,658.00& 1,654.30& 1,656.307& 1,657.50& 1,657.50& 1,656.72\\
BM3D  &  3.17&	1.51&	3.62&	5.15&	4.39&	3.57\\
PID  & 2,184.10&	2,147.30&	2,091.40&	2,064.80&	2,074.30&	2,112.38\\
KSVD  & 340.96&	332.40&	327.14&	323.47&	304.35&	325.66\\
NCSR & 1,201.80&	1,565.90&	1,125.60&	935.60&	947.97&	1,155.37\\
OCTOBOS &1,444.40&	2,326.80&	1,425.70&	927.39&	984.72&	1,421.80\\[2pt]
\hline
\end{tabular}
\end{center}
\end{table}
\subsubsection{Memory Consumption.} 
\begin{figure}
\centering
\includegraphics[width=9cm]{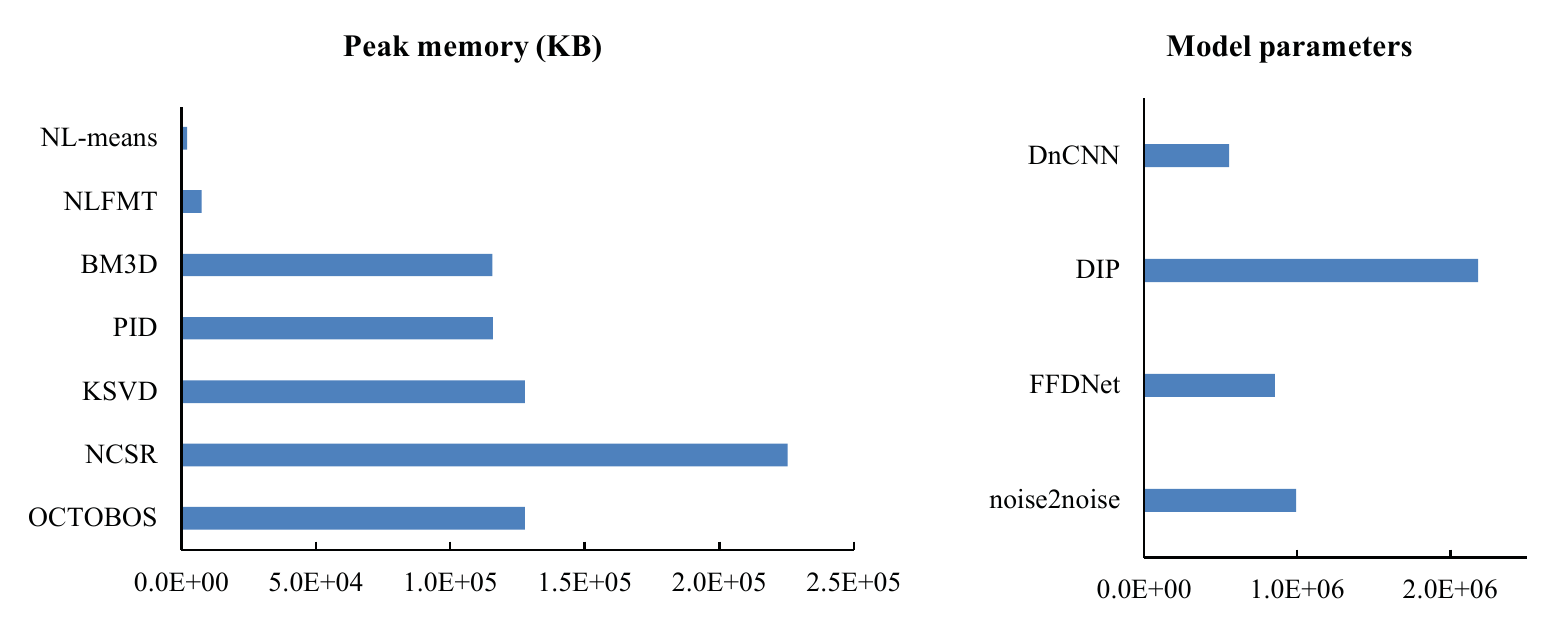}
\caption{The peak memory (left) and parameter number (right) of denoising approaches.}
\label{fig:memory}
\end{figure}

We observe the peak memory occupied by the filter-based approaches and sparse representation approaches in image denoising (see Fig.\ref{fig:memory}). NL-means and NLFMT use the least amount of memory, and the rest approaches generally take up more than 100,000 KB memory. For the deep learning approaches, we use another indicator to reflect the memory consumption, that is, to compare the number of parameters in these networks, show in Fig.\ref{fig:memory}. Because we consider packaging the trained model (except DIP) directly into the satellites, the model with fewer parameters takes up a relatively small amount of memory space.

On the whole, the filter-based approaches can achieve high image denoising quality in general, especially the BM3D is excellent in every indicator, the sparse representation approaches are normal on each indicator, and the deep learning approaches have some approaches achieving high image denoising quality, and have a great advantage in time cost and memory consumption.

\section{Strategies for Image Denoising on Satellite TianZhi-1}
After analyzing the excellent image denoising approaches on the ground in recent years, we propose two feasible strategies for image denoising on satellites based on TianZhi-1. TianZhi-1 is a software-defined satellite, is the first technical verification satellite of the "TianZhi" series led by Institute of Software Chinese Academy of Science, which was launched on November 20, 2018. Its main payloads include one small cloud computing platform and four reinforced Chinese-made smart-phones.

Based on the comparison of simulation results, BM3D and FFDNet are worthy of further research and have the potential to be applied to TianZhi-1. We have two options: 1) BM3D on cloud computing platform and 2) FFDNet application on smart-phone. BM3D is a traditional algorithm that does not require clean images for training, has advantages in the absence of clean remote sensing images. However, its operation requires over one hundred megabytes of memory, it is more difficult to run on a smart-phone. We can further streamline the algorithm by cutting off superfluous functions and optimize functions memory calls to satisfy the cloud computing platform on TianZhi-1. On the other side, FFDNet network structure is small, the image denoising quality is outstanding, especially in the case of high proportion noises, which is better than BM3D, moreover, the denoising time of the trained model on the CPU takes only a few seconds (about 5 seconds), and the occupied memory is about ten megabytes. So we can further simplify the FFDNet network and carry out transfer learning to compensate for the lake of clean remote sensing images, then package the trained FFDnet model into an application, test it on the same smart-phone of TianZhi-1 on the ground, upload the tested application to the smart-phone on TianZhi-1.

\section{Conclusion}
In this paper, we first summarize different image denoising approaches, then carry out simulation experiments on remote sensing images based on the source code provided, compared on the four indicators: PSNR, SSIM, time cost and memory consumption, all of them can determine whether an approach is suitable for image denoising on satellites. At last, from the simulation results, we propose two strategies based on BM3D and FFDNet that have the potential to be applied to satellites based on TianZhi-1.


%
%


\begin{thebibliography}{20}
%

\bibitem {bua:col:mor}
Buades, A., Coll, B., Morel, JM.: A non-local algorithm for image denoising. In Computer Vision and Pattern Recognition, 2005. CVPR 2005. IEEE Computer Society Conference on, vol. 2, pp. 60-65 (2005).
\bibitem {kum}
Kumar, B.S.: Image denoising based on non-local means filter and its method noise thresholding. Signal, image and video processing 7, no. 6, 1211-1227 (2013).
\bibitem {dab:foi:kat:egi}
Dabov, K., Foi, A., Katkovnik, V., Egiazarian, K.: Image denoising by sparse 3-D trans-form-domain collaborative filtering. IEEE Transactions on image processing 16, no. 8, 2080-2095 (2007).
\bibitem {kna:zwi:pid}
Knaus, C., Zwicker, M.: Progressive image denoising. IEEE transactions on image processing 23, no. 7, 3114-3125 (2014).
\bibitem {kna:zwi:ddid}
Knaus, C., Zwicker, M.: Dual-domain image denoising. In Image Processing (ICIP), 2013 20th IEEE International Conference on, pp. 440-444. (2013).
\bibitem {ela:aha}
Elad, M., Aharon, M.: Image denoising via sparse and redundant representations over learned dictionaries.IEEE Transactions on Image processing 15, no. 12, 3736-3745 (2006).
\bibitem {don:zha:shi:li}
Dong, W., Zhang, L., Shi, G., Li, X.: Nonlocally centralized sparse representation for image restoration. IEEE Transactions on Image Processing 22, no. 4, 1620-1630 (2013).
\bibitem {wen:rav:bre}
Wen, B., Ravishankar, S., Bresler, Y.: Structured overcomplete sparsifying transform learning with convergence guarantees and applications. International Journal of Computer Vision 114, no. 2-3, 137-167 (2015).
\bibitem {rav:bre}
Ravishankar, S., Bresler, Y.: MR image reconstruction from highly undersampled k-space data by dictionary learning.IEEE transactions on medical imaging 30, no. 5, 1028 (2011).
\bibitem {jai:seu}
Jain V, Seung S.: Natural image denoising with convolutional networks. In Advances in Neural Information Processing Systems, pp. 769-776. (2009).
\bibitem {bur:sch:har}
Burger, H.C., Schuler, C.J., Harmeling S.: Image denoising: Can plain neural networks compete with BM3D?. In Computer Vision and Pattern Recognition (CVPR), 2012 IEEE Conference on, pp. 2392-2399. (2012).
\bibitem {sch:rot}
Schmidt, U., Roth, S.: Shrinkage fields for effective image restoration. In Proceedings of the IEEE Conference on Computer Vision and Pattern Recognition, pp. 2774-2781. (2014).
\bibitem {che:poc}
Chen, Y., Pock, T.: Trainable nonlinear reaction diffusion: A flexible framework for fast and effective image restoration. IEEE transactions on pattern analysis and machine intelligence 39, no. 6, 1256-1272 (2017).
\bibitem {mao:she:yan}
Mao, X., Shen, C., Yang, Y.B.: Image restoration using very deep convolutional en-coder-decoder networks with symmetric skip connections. In Advances in neural information processing systems, pp. 2802-2810. (2016).
\bibitem {zha:zuo:che:men:zha}
Zhang, K., Zuo, W., Chen, Y., Meng, D., Zhang, L.: Beyond a gaussian denoiser: Resid-ual learning of deep cnn for image denoising. IEEE Transactions on Image Processing 26, no. 7, 3142-3155 (2017).
\bibitem {he:zha:ren:sun}
He, K., Zhang, X., Ren, S., Sun, J.: Deep residual learning for image recognition. In Proceedings of the IEEE conference on computer vision and pattern recognition, pp. 770-778. (2016).
\bibitem {uly:ved:lem}
Ulyanov, D., Vedaldi, A., Lempitsky, V.: Deep image prior. arXiv preprint arXiv:1711-10925 (2017).
\bibitem {zha:zuo:zha}
Zhang, K., Zuo, W., Zhang, L.: FFDNet: Toward a fast and flexible solution for CNN based image denoising. IEEE Transactions on Image Processing (2018).
\bibitem {leh:mun:has:lai:kar:ait:ail}
Lehtinen, J., Munkberg, J., Hasselgren, J., Laine, S., Karras, T., Aittala, M., Aila, T.: Noise2Noise: Learning Image Restoration without Clean Data. arXiv preprint arXiv:1803.04189 (2018).
\bibitem {mag:tar:cha:all}
Maggiori, E., Tarabalka, Y., Charpiat, G., Alliez, P.: Can semantic labeling methods generalize to any city? the inria aerial image labeling benchmark. InIEEE Interna-tional Symposium on Geoscience and Remote Sensing (IGARSS) (2017).
\end{thebibliography}
\end{document}